# Matrix-aware 2D field-effect biosensing of cardiac troponin I using complementary ion-sensitive transduction

Subhabrata Das[a], Bharti[b], Babulu Pradhan[c], Abhishek R Nath[a], Kashish Sapra[d], Jaydeep sharma[a], Vijayender Bhalla[b], Satyaprakash Sahoo[c], Manish Mathew[d], Md. Ehesan Ali[a], Kaushik Ghosh[a]*

[a]Quantum Materials & Devices Unit, Institute of Nano Science and Technology, Knowledge City-Sector 81, Mohali 140306, India.

[b]Council of Scientific and Industrial Research (CSIR), Institute of Microbial Technology (IMTECH), Sector 39-A, Chandigarh, 160036, India.

[c]Laboratory for Low-Dimensional Materials, Institute of Physics, Bhubaneswar-751005, India

[d]CSIR-Central Electronics Engineering Research Institute, Pilani, Rajasthan, 333031, India

**Corresponding Author:** **kaushik@inst.ac.in**

## Abstract

Cardiovascular diseases remain a leading cause of mortality worldwide, highlighting the need for rapid, sensitive and reliable detection of cardiac injury at the point of care. Here, we develop a dual-sensor platform that combines an aptamer-functionalized molybdenum disulfide ($MoS_2$) field-effect transistor (2D-FET) with a graphene-based ion-sensitive field-effect transistor (ISFET) for the detection of cardiac troponin I (Trop-I) in complex biological matrices. The atomically thin $MoS_2$ channel enables highly sensitive detection of Trop-I, achieving a limit of detection of 0.001 ng $mL^{-1}$. However, the high ionic strength of serum substantially attenuates the field-effect response, resulting in approximately 80% degradation of the sensor response. To address this matrix effect, the ISFET is employed as a secondary sensor to characterize the ionic environment of the sample. Machine-learning-assisted integration of the 2D-FET and ISFET responses is investigated as a strategy for compensating serum-induced signal distortion and recovering key electrical figures of merit extracted from the 2D-FET transfer characteristics, including maximum on-state current ($I_{ds}^{max}$), subthreshold slope (SS) and transconductance ($g_m$). This dual-sensor architecture combines molecular recognition, high analytical sensitivity and matrix-aware electrical signal correction, providing a potential pathway towards more reliable FET-based biomarker detection in complex biological fluids. The approach may further enable multiplexed detection of cardiac biomarkers through incorporation of target-specific aptamers, offering potential for broader assessment of cardiac stress and myocardial injury risk.

## Introduction

Cardiovascular diseases (CVDs) remain the leading cause of mortality worldwide, representing a major global health burden and accounting for millions of deaths annually[1]. The clinical assessment of cardiovascular disorders encompasses a broad range of diagnostic approaches, from cardiovascular risk evaluation and assessment of cardiac function to the timely identification of acute myocardial injury and myocardial infarction. Accordingly, clinical evaluation relies on an integrated combination of physiological assessment, electrocardiography, echocardiography, cardiac imaging, and circulating biochemical biomarkers[2]. Among these biomarkers, cardiac troponin I (cTnI) has emerged as a cardiac-specific indicator of myocardial injury and, together with cardiac troponin T, plays a central role in the assessment of patients with suspected acute coronary syndromes[2,3]. Conventional clinical quantification of cTnI predominantly relies on high-sensitivity immunoassays, which provide excellent analytical sensitivity and precision but are commonly implemented using automated laboratory-based instrumentation[4]. Although high-sensitivity point-of-care troponin assays are emerging, their clinical implementation requires appropriate analytical validation, quality control, and integration into diagnostic workflows[5]. These considerations have stimulated increasing interest in portable biosensing platforms capable of rapid, sensitive, and selective biomarker analysis. Nevertheless, translating the analytical performance of emerging biosensors from controlled laboratory conditions to complex physiological matrices remains a significant challenge, highlighting the need for sensing platforms that combine high sensitivity with reliable operation in real biological samples[6,7].

The development of portable biosensors has increasingly relied on engineered nanomaterials to enhance the interaction between target biomolecules and transducer surfaces. Noble metals, metal oxides, carbon-based materials, metal–organic frameworks (MOFs) and their composites have been extensively investigated because of their high surface area, tunable surface chemistry and favourable physicochemical properties for biomolecular immobilization[8–10]. However, achieving both efficient molecular recognition and sensitive, reproducible electrical transduction remains challenging, particularly when the recognition event occurs at the interface of a nanoscale sensing material. Two-dimensional (2D) materials have emerged as particularly attractive transducers because of their atomically thin structures, high surface accessibility and pronounced sensitivity to changes in their local chemical and electrostatic environment[6,11]. In contrast to bulk materials, a substantial fraction of the atoms in a 2D layer are exposed to the surrounding medium, allowing surface-bound recognition elements to efficiently modulate the electronic properties of the underlying material. These characteristics are especially advantageous for aptamer-based sensing, in which target binding at the sensor interface can induce changes in local charge distribution and thereby generate a measurable electrical response[6,12].

Beyond the choice of sensing material, the transducer architecture critically determines the efficiency of converting molecular recognition into an electrical signal. Interdigitated-electrode (IDE) resistive sensors primarily rely on analyte-induced changes in the conductance of the sensing layer[13], whereas field-effect transistors (FETs) exploit electrostatic modulation of a semiconducting channel to transduce interfacial molecular interactions[14,15]. In a FET biosensor, target binding can perturb the local surface potential and carrier density, producing amplified changes in the drain current and transfer characteristics[12,14,15]. This effect is particularly pronounced in 2D-FETs, where the atomically thin channel provides strong electrostatic coupling between surface-bound recognition elements and the conducting channel[6,11,16,17]. Consequently, 2D-FETs offer highly sensitive, label-free electrical transduction together with multiple device-level figures of merit, including maximum on-state current, subthreshold slope and transconductance, enabling more comprehensive characterization of biomolecular interactions than a single resistive response[12,14,18]. However, the strong interfacial electrostatic coupling that underpins this sensitivity also renders 2D-FETs susceptible to ionic screening in physiological fluids, posing a fundamental challenge to their reliable operation in complex biological matrices[18–21].

Physiological fluids such as serum present a highly ionic environment that can substantially influence interfacial electrostatic sensing. Electrolytes including $Na^+$, $K^+$, $Cl^-$ and $Ca^{2+}$ contribute to the ionic strength of biological fluids and play essential roles in maintaining physiological homeostasis and cellular signaling[22]. The resulting ionic screening can attenuate the electrostatic coupling between surface-bound recognition events and the semiconductor channel of a FET, thereby limiting direct electrical sensing in physiological matrices[19,20]. Rather than being considered solely as an interfering component, the ionic environment can therefore serve as a measurable descriptor of the sample matrix. Ion-sensitive field-effect transistors (ISFETs) provide an established electrical approach for probing ionic and proton activity in liquid environments[23]. ISFET-based systems have been demonstrated for the determination of $Na^+$ and $K^+$ in blood and serum[24,25], while ion-selective FET sensor arrays have also enabled the analysis of $Na^+$, $K^+$ and $Cl^-$ in blood serum[26,27]. Furthermore, ISFET platforms have been employed for pH monitoring[28] in human interstitial fluid and serum-based interstitial-fluid-mimicking samples, as well as for real-time monitoring of extracellular acidification associated with cellular metabolic activity[29].

These studies establish the capability of ISFETs to sensitively characterize ionic conditions in physiologically relevant environments. However, their potential as a complementary matrix-sensing element for improving the reliability of a separate biomolecular FET sensor remains largely unexplored. Such an approach could provide additional information on the ionic environment surrounding the primary sensing interface, enabling matrix-dependent variations in the biomolecular FET response to be distinguished and subsequently compensated.

Here, we develop a dual-sensor platform for sensitive and reliable detection of cTnI in complex biological matrices. An aptamer-functionalized $MoS_2$ 2D-FET is employed as the primary biosensor for selective cTnI detection, while device architectures based on interdigitated electrodes are comparatively investigated to establish the sensitivity advantage of field-effect transduction. To address the substantial degradation of the 2D-FET response in Trop-I-spiked serum arising from the high ionic strength of the physiological matrix, a graphene-based ISFET is introduced as a secondary sensor to characterize the ionic environment of the sample. The electrical responses of the primary 2D-FET and secondary ISFET are subsequently integrated through machine-learning approaches to account for matrix-induced variations in the FET response and to investigate the recovery of key transfer-characteristic figures of merit, including maximum on-state current ($I_{ON}$), subthreshold slope (SS) and transconductance ($g_m$). This complementary sensing strategy establishes a matrix-aware framework for improving the reliability of 2D-FET-based biomarker detection and provides a potential pathway towards portable and multiplexed cardiovascular biomarker sensing.

## Synthesis and Structural Characterization of Basal-Plane Monolayer $MoS_2$

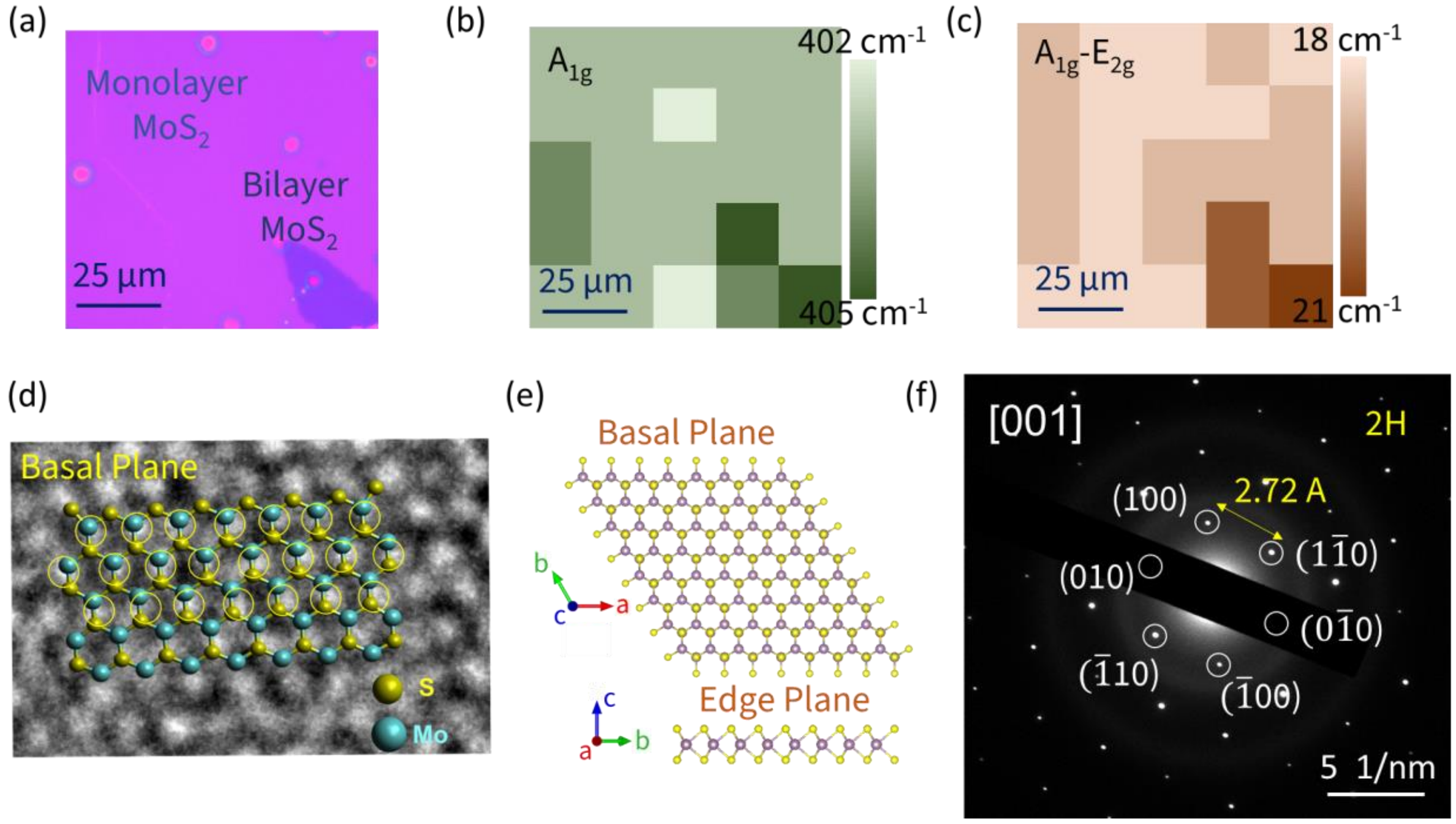

**Figure 1 Structural characterization and basal-plane identification of monolayer $MoS_2$.** (a) Optical image showing monolayer and bilayer $MoS_2$ domains. (b,c) Raman maps of the $A_{1g}$ mode and the separation between the $A_{1g}$ and $E_{2g}$ modes, respectively, showing values in the ranges of 402–405 and 18–21 $cm^{-1}$. (d) HRTEM image revealing the basal-plane structure of $MoS_2$. (e) Schematic illustration distinguishing the basal and edge planes of $MoS_2$. (f) Selected-area electron diffraction (SAED) pattern confirming the 2H phase of $MoS_2$.

$MoS_2$ was grown by atmospheric-pressure chemical vapour deposition (APCVD) on 1.2 × 1.2 $cm^2$ Si/$SiO_2$ substrates with a 300-nm-thick $SiO_2$ layer. Molybdenum trioxide ($MoO_3$) and sulfur (S) powders were used as the Mo and S precursors, respectively, with a precursor ratio of 1:100 under an Ar carrier-gas environment at a growth temperature of approximately 750 °C. The detailed growth procedure is provided in **Supplementary Information 1**. Two monolayer $MoS_2$ configurations were employed for the sensing platforms: continuous monolayer films for resistive sensing and individual triangular monolayer domains for transistor devices. Characterization of the continuous film is presented in **Fig. 1**, whereas the structural and optical characterization of the triangular monolayer domains is provided in **Supplementary Information 2**. **Figure 1a** shows the optical contrast between monolayer and bilayer $MoS_2$ regions within the grown film. Raman mapping further distinguishes the two thicknesses. The spatial distribution of the $A_{1g}$ mode (**Fig. 1b**) exhibits Raman shifts of approximately 402–403 $cm^{-1}$ for monolayer regions and 404–405 $cm^{-1}$ for bilayer regions. The separation between the $A_{1g}$ and $E_{2g}$ modes (Δω) provides an additional thickness-sensitive parameter, with values of approximately 18.5–20 $cm^{-1}$ for the monolayer and >20 $cm^{-1}$ for the bilayer regions (**Fig. 1c**), consistent with the established Raman characteristics of monolayer and bilayer $MoS_2$[30,31]. The monolayer nature of the triangular $MoS_2$ domains was further confirmed by tapping-mode atomic force microscopy (AFM), which yielded a thickness of approximately 0.65 nm (Supplementary Fig. 2), consistent with previously reported AFM measurements of monolayer $MoS_2$. The corresponding photoluminescence peak at approximately 1.84 eV is characteristic of the direct excitonic transition of monolayer $MoS_2$ and supports its semiconducting character[32,33] (**Supplementary Information 2**). The exposed basal plane is particularly relevant to the subsequent aptamer-functionalization strategy because it constitutes the dominant surface area of the atomically thin $MoS_2$ channel. HRTEM resolves the atomic arrangement of the basal plane (**Fig. 1d**), while **Fig. 1e** schematically distinguishes the basal and edge planes[34]. The TEM sample-preparation procedure is provided in **Supplementary Information 3**. The crystallinity of the $MoS_2$ was further examined by selected-area electron diffraction (SAED), which shows well-defined diffraction spots corresponding to the hexagonal $MoS_2$ lattice (**Fig. 1f**). Together with the characteristic Raman and PL responses, these structural observations are consistent with highly crystalline, semiconducting 2H-$MoS_2$[31,35]. Overall, the grown $MoS_2$ exhibits the structural and optical characteristics required for the subsequent sensing studies, including monolayer thickness, a crystalline semiconducting phase and an accessible basal-plane surface for aptamer functionalization.

## Defect-Assisted Aptamer Immobilization and Binding on $MoS_2$

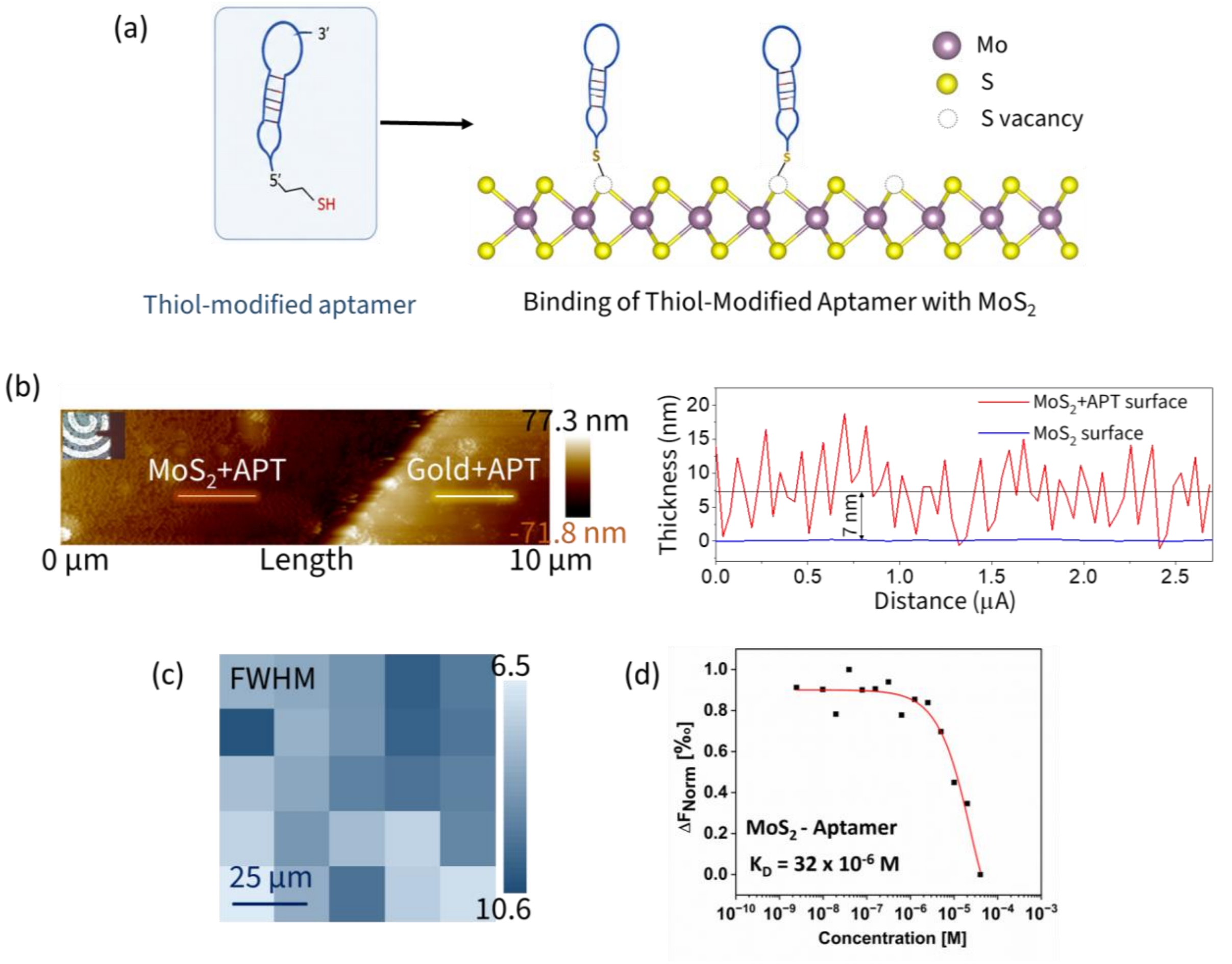

**Figure 2 Aptamer functionalization and binding characterization of the $MoS_2$ surface.** (a) Schematic illustration of thiol-modified aptamer immobilization on the $MoS_2$ basal plane through interaction with reactive defect-associated surface sites. (b) AFM topographic image and corresponding height profiles of the $MoS_2$+APT region and Au+APT reference surface. (c) Spatial map of the $A_{1g}$ Raman FWHM across the $MoS_2$ surface. (d) MST binding curve obtained by titrating 5′-thiol- and 3′-FAM-labelled aptamer with increasing $MoS_2$ concentration, yielding an apparent $K_D$ of 32 X $10^{-6}$ M.

To establish the interaction between the thiol-modified aptamer and the $MoS_2$ surface, a combination of structural, morphological and binding analyses was performed. The proposed immobilization mechanism is illustrated in **Fig. 2a**. The aptamer was functionalized with a terminal thiol (-SH) group, providing a direct anchoring functionality for interaction with reactive sites on the $MoS_2$ surface without requiring an additional molecular linker. Although the basal plane of 2H-$MoS_2$ is relatively inert, intrinsic lattice defects, particularly sulfur vacancies, can generate coordinatively unsaturated Mo sites that are more chemically reactive. Such defect-associated sites provide a plausible basis for thiol-mediated interaction between the aptamer and the $MoS_2$ basal plane[36–38]. The defect characteristics of the as-grown $MoS_2$ were first examined using spatially resolved Raman spectroscopy. The Raman map of the $A_{1g}$ mode exhibited relatively small spatial variation in peak position, indicating a comparatively uniform vibrational environment across the investigated region (**Fig. 2c**). In contrast, the corresponding $A_{1g}$ linewidth showed a broader distribution, with FWHM values ranging from approximately 6.5 to 10.6 $cm^{-1}$. Such linewidth broadening is associated with enhanced phonon scattering and can reflect variations in structural disorder and defect-related perturbations in $MoS_2$[39]. Notably, the larger FWHM values were more pronounced in the monolayer region than in the bilayer region, indicating greater defect-related broadening in the exposed monolayer layer. Although the Raman linewidth alone does not uniquely identify sulfur vacancies, the observed defect-related broadening is consistent with the presence of reactive defect sites that can contribute to thiol functionalization of the $MoS_2$ basal plane. The surface interaction of the aptamer with $MoS_2$ was subsequently examined by atomic force microscopy (AFM). The AFM image in **Fig. 2b** shows distinct surface features in the $MoS_2$+APT region compared with the reference Au+APT region, while the corresponding height profiles reveal additional nanoscale surface features associated with the $MoS_2$-functionalized region. These changes in surface morphology are consistent with the presence of an aptamer-containing layer on the $MoS_2$ surface. The AFM measurement strategy and the corresponding comparison of aptamer distribution on the Au and $MoS_2$ regions are further presented in **Supplementary Fig. 4**. In particular, the pronounced surface features observed on the Au+APT region suggest local aptamer aggregation, whereas the $MoS_2$+APT region exhibits a comparatively distributed surface response following functionalization. To quantitatively investigate the $MoS_2$–aptamer interaction, microscale thermophoresis (MST) was subsequently performed (**Fig. 2d**). A constant concentration of FAM-labelled aptamer was titrated against increasing concentrations of $MoS_2$, resulting in a concentration-dependent change in the thermophoretic response. Fitting of the binding curve yielded an apparent dissociation constant ($K_D$) of 32 X $10^{-6}$ M, demonstrating a measurable interaction between the aptamer and $MoS_2$. Because the interaction occurs at a heterogeneous two-dimensional surface rather than at a single well-defined molecular binding site, the measured value is considered an apparent $K_D$[40]. Taken together, the defect-sensitive Raman response, AFM observations and MST measurements support the interaction and immobilization of the thiol-functionalized aptamer on the $MoS_2$ surface[36–38]. However, the precise chemical nature of thiol attachment to 2H-$MoS_2$ remains debated in the literature, with alternative studies reporting thiol oxidation and subsequent physisorption rather than direct covalent attachment[41,42]. Thus, the proposed vacancy-assisted interaction provides a plausible molecular interpretation of the observed aptamer association and supports the subsequent Trop-I recognition process at the $MoS_2$ interface.

**Optimization of Interdigitated-Electrode Geometry for Trop-I Detection**

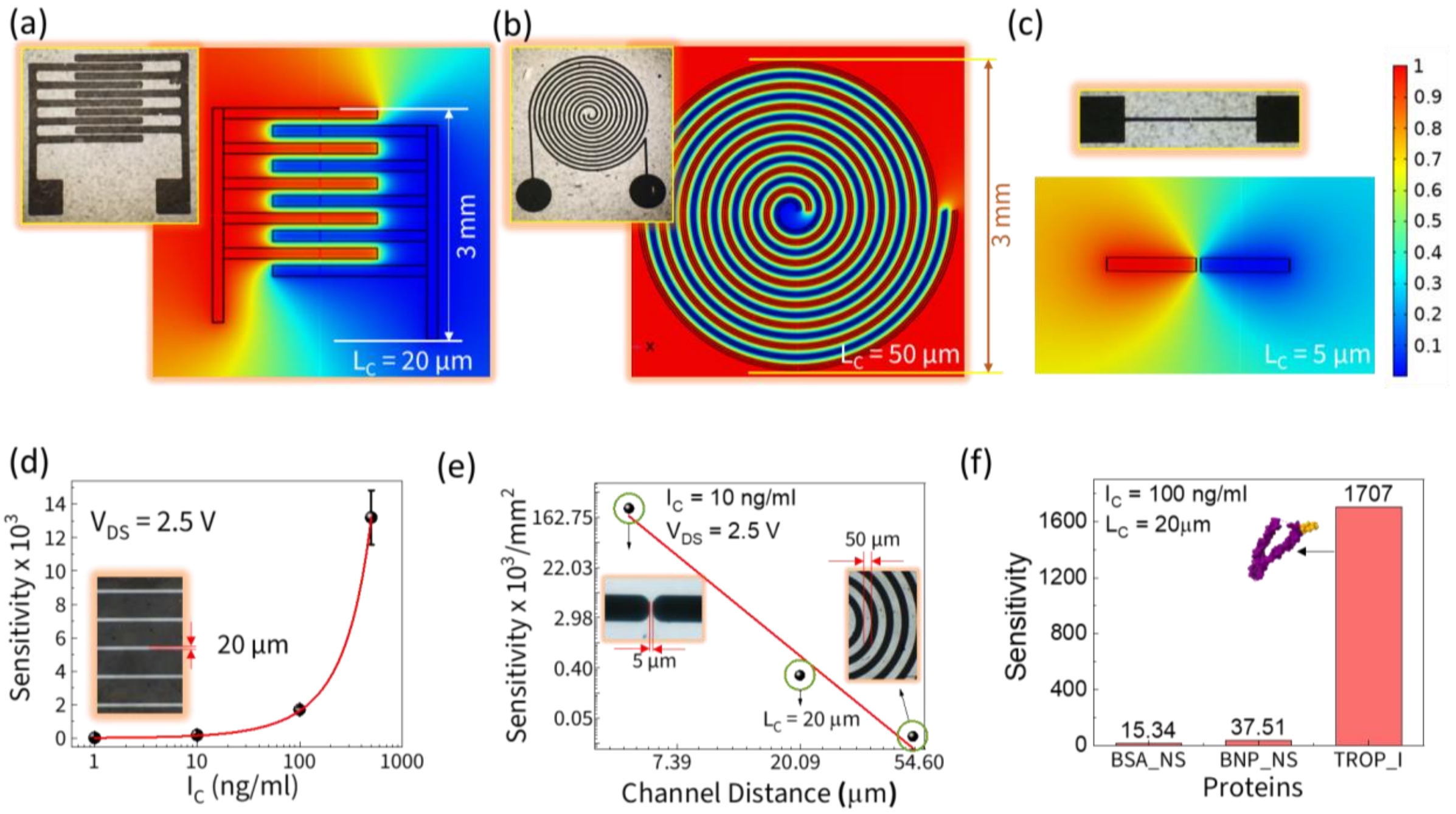

**Figure 3 Electrical response of Trop-I sensing devices with different interdigitated-electrode geometries and channel lengths.** (a) Finger-pattern IDE with a 20-µm channel length, (b) coil-pattern IDE with a 50-µm channel length, and (c) arm-pattern IDE with a 5-µm channel length, together with the corresponding potential distributions, fabricated on continuous monolayer $MoS_2$. (d) Concentration-dependent Trop-I response of the 20-µm-channel device measured at $V_{DS}$ = 2.5 V. (e) Areal sensitivity as a function of channel length. (f) Selective response of the Trop-I sensor compared with non-specific responses to bovine serum albumin (BSA) and brain natriuretic peptide (BNP).

Following aptamer immobilization, the Trop-I response of continuous monolayer $MoS_2$ was first evaluated in the resistive configuration using interdigitated electrodes (IDEs) with different geometries and channel lengths (**Fig. 3a–c**). The devices were compared to examine the influence of electrode geometry, active sensing area and channel length on the electrical response[43–46]. At 10 ng mL$^{-1}$ Trop-I, the finger- and coil-pattern IDEs, both defined within a 3-mm lateral dimension, exhibited total sensitivities of 72.33 and 93.51, respectively (**Supplementary information 5**). The higher absolute response of the coil geometry is accompanied by a substantially larger active area (3.53 mm$^2$ versus 0.24 mm$^2$ for the finger geometry). However, normalization to the active area reveals markedly different area-specific responses, with the finger device exhibiting a sensitivity of 301.38 mm$^{-2}$ compared with 26.49 mm$^{-2}$ for the coil device. The arm-pattern IDE, with a 5-µm channel length and an active area of 0.25 X $10^{-3}$ mm$^2$, exhibits the highest normalized sensitivity of 2.32 X $10^5$ mm$^{-2}$ although its total sensitivity remains lower at 58. These results indicate that the electrical response is governed by the combined effects of sensing area and electrode spacing rather than by active area alone[43–45]. The concentration-dependent response of the 20-µm-channel device, measured at $V_{DS}$ = 2.5 V, increased with Trop-I concentration and could be described by an exponential dependence (**Fig. 3d**). A measurable response was obtained down to 1 ng $mL^{-1}$; however, further reduction in channel length or increase in active area did not provide a sufficient response in the sub-ng $mL^{-1}$ regime investigated (**Fig. 3e**). This limitation is important for translation to clinically relevant cTnI measurements, for which decision thresholds are assay-specific and are commonly expressed in the low-ng $L^{-1}$ range rather than through a universal concentration threshold[2]. The selectivity of the aptamer-functionalized $MoS_2$ sensor was subsequently evaluated against BSA and BNP as non-target proteins (**Fig. 3f**). At an identical concentration of 100 ng $mL^{-1}$, the responses to BSA and BNP were 15.34 and 37.51, respectively, compared with 1717 for Trop-I, demonstrating a substantially higher response towards the target analyte under the same measurement conditions. These results confirm the selective recognition capability of the immobilized aptamer and establish the suitability of the $MoS_2$ surface for Trop-I sensing[47,48]. Nevertheless, the limited response of the resistive architecture at low Trop-I concentrations motivated the transition to a three-terminal field-effect configuration, where surface binding can be coupled directly to electrostatic modulation of the $MoS_2$ channel.

**$MoS_2$ transistor to amplify the Trop-I signal**

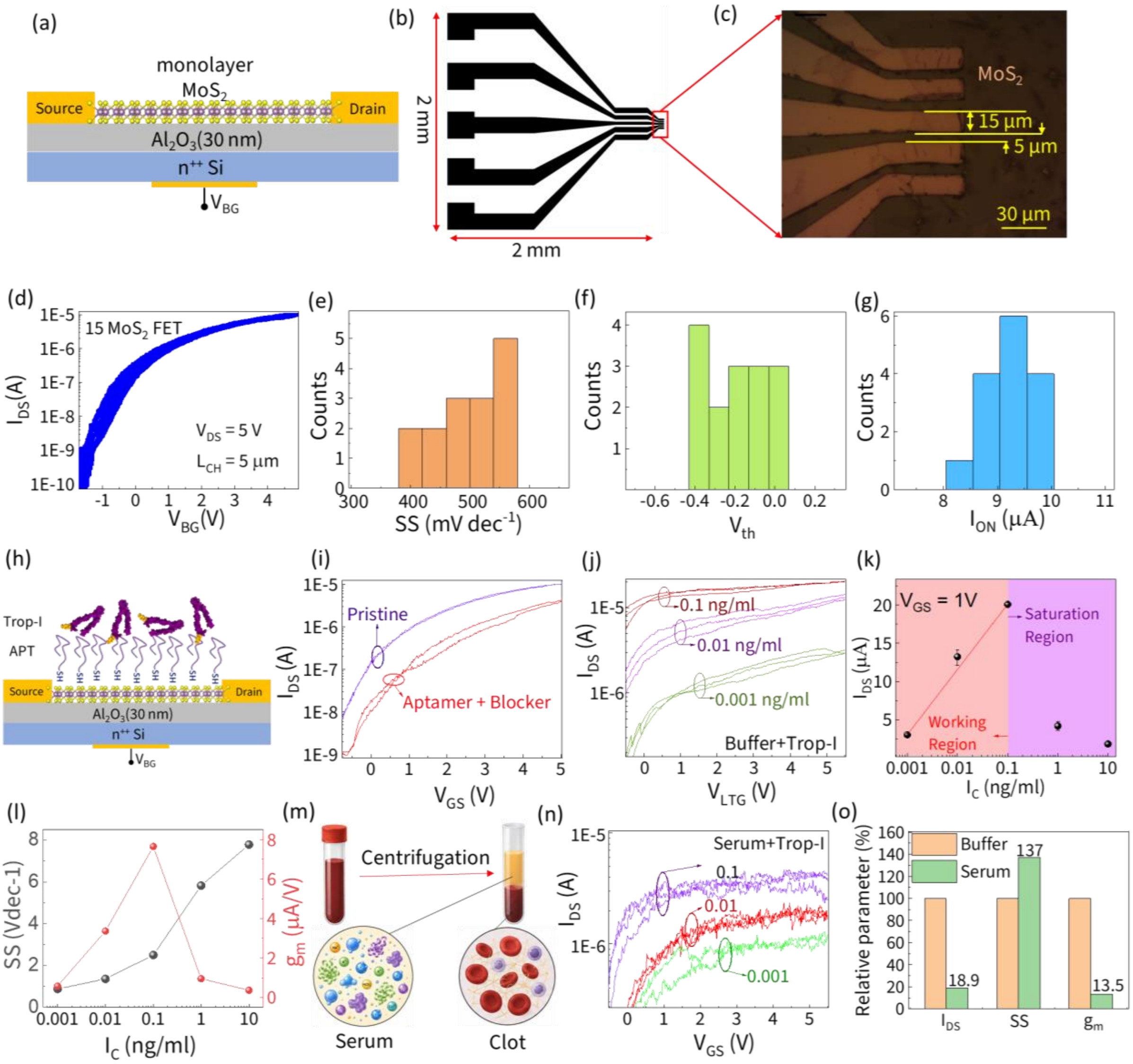

**Figure 4 Electrical characterization of the monolayer $MoS_2$ FET and its application to cardiac Troponin-I (Trop-I) detection in buffer and serum.** (a) Schematic cross-section of the back-gated $MoS_2$ transistor incorporating a 30 nm $Al_2O_3$ gate dielectric on an $n^{++}$-Si substrate. (b) Schematic layout of the transistor device array and electrode geometry. (c) Optical micrograph of the fabricated $MoS_2$ FET showing the channel and electrode dimensions. (d) Transfer characteristics of 15 $MoS_2$ FETs measured at $V_{DS}$=5 V. (e–g) Statistical distributions of subthreshold swing (SS), threshold voltage ($V_{TH}$), and on-state current ($I_{ON}$), demonstrating device-to-device uniformity. (h) Schematic illustration of Trop-I-specific aptamer functionalization followed by BSA blocking. (i) Transfer characteristics of the pristine and aptamer-functionalized, BSA-blocked $MoS_2$ FET. (j) Transfer characteristics measured at different Trop-I concentrations in buffer. (k) Concentration-dependent drain current at ($V_{GS}$) V, identifying the high-response and response-decline regimes. (l) Variation of SS and $g_m$ with Trop-I concentration. (m) Schematic illustration of serum preparation by centrifugation to separate the serum from the clot. (n) Transfer characteristics of the $MoS_2$ FET measured with Trop-I-spiked serum samples at different concentrations. (o) Comparison of the relative $I_{ON}$, SS, and $g_m$ in buffer and serum, highlighting the influence of the serum matrix on transistor performance.

Three terminal device gate mediated channel carrier density modulation give access to the enhancement of the sensitivity[12,16]. Also multiple figure of merit (FoM) can be extracted from the transfer characteristic which associated with the different interectional information carrier. The cross-sectional schematic of $MoS_2$ thin film transistor (TFT) is shown in **Fig 4a** and the device statistical stability and device based sensing response is shown in **Fig 4**. For conducting back gate $n^{++}$ silicon is taken for back gating as well as substrate, high dielectric constant (k) 30 nm oxide material $Al_2O_3$ deposited via ALD process for better controbility over channel material ($MoS_2$) conduction[49]. For higher mobility the source and drain are made in a single triangular flake the electrodes arms are extend to minimize the effect of the all analytic interection with contact (**Fig. 4b**). **Fig. 4c** shows the microscopic image of the $MoS_2$ transistor having channel distance of 5 micron. 15 $MoS_2$ transistor transfer characteristic shown in **Fig. 4d** where at a constant 5 volt of the drain source voltage ($V_{DS}$) the gate voltage ($V_{BG}$) sweep in the range of -1.5 to 5 volt. Device stability is presented via statistical analysis (**Fig. 4e-g**) of the parameters subthreshold slope (SS), threshold voltage ($V_{th}$) and on current ($I_{ON}$)[50]. Subthreshold slope signifies the responsiveness of tunable channel, lower the value higher the controbility. The extracted threshold voltages ($V_{TH}$) ranged from −0.4 to 0.05 V, indicating the operating characteristics of the fabricated $MoS_2$ FETs. Such variations can be influenced by several factors, including substrate-induced doping, fixed charges within the $Al_2O_3$ dielectric, and interfacial effects. $I_{ON}$ reached to the level of 8 to 10 μA from the off current ($I_{OFF}$) 0.1 nA, and make the 5 order of current enhacement that make the device applicable for higher sensing detectable platform[51,52].

The $MoS_2$ FET channel was functionalized with a Trop-I-specific aptamer, followed by bovine serum albumin (BSA) blocking to minimize nonspecific adsorption[53], as illustrated in **Fig. 4h**. The transfer characteristics exhibited a reduction in drain current following functionalization compared with the pristine device (**Fig. 4i**). This change can be attributed to modifications in the local electrostatic environment and charge-transport properties of the $MoS_2$ channel. In particular, the negatively charged phosphate backbone of the immobilized aptamer can influence electron accumulation in n-type $MoS_2$, while the additional biomolecular layer introduced by BSA may further modify the interfacial environment. Biomolecular coverage near the source/drain contacts may also influence charge injection. These effects collectively provide a plausible explanation for the reduced drain current following functionalization. The concentration-dependent sensing response was subsequently investigated by exposing the functionalized transistor to different Trop-I concentrations and extracting three figures of merit-on-state current ($I_{ON}$), subthreshold swing (SS), and transconductance ($g_m$) - from the transfer characteristics (**Fig. 4j–l**). At lower Trop-I concentrations, specific aptamer–analyte binding modifies the local electrostatic potential, resulting in increased drain current and transconductance[52]. Beyond approximately 0.1 ng $mL^{-1}$, both parameters decrease, whereas SS increases continuously. This behaviour may be associated with progressive changes in the interfacial charge distribution, effective gate coupling, and potential charge-trapping effects as the biomolecular layer evolves with increasing analyte concentration. The increasing SS indicates less efficient subthreshold switching, while the decrease in $g_m$ effects weaker gate-induced current modulation at the extraction point. The reduction in $I_{ON}$ may additionally arise from changes in the transfer-curve position relative to the selected operating voltage. Together, these correlated changes suggest a transition from an initial high-response regime to a response-decline regime at higher Trop-I concentrations[54–56], highlighting the value of multiple transistor figures of merit for evaluating the sensing response.

The applicability of the developed sensor under physiologically relevant conditions was subsequently investigated using human blood-derived samples. Following sample preparation, Trop-I was spiked into the serum and measured using the same aptamer-functionalized $MoS_2$ FET platform (**Fig. 4n**). Compared with measurements in the buffer medium, the serum matrix produced a substantial deterioration in the transistor response, with reductions of more than 80% in both $I_{ON}$ and $g_m$, accompanied by an increase in SS (**Fig. 4o**). These changes suggest that the complex ionic environment of serum modifies the effective electrostatic coupling between the biomolecular recognition interface and the $MoS_2$ channel. In particular, ionic screening can attenuate the electrostatic influence of the bound Trop-I molecules[12,19–21,57], while matrix-induced interfacial effects may further alter the gate-controlled transport behaviour[12]. Consequently, the electrical response of the FET reflects the combined contributions of Trop-I binding and the surrounding physiological matrix, complicating the reliable determination of the biomarker concentration using a single sensing channel. To address this challenge, a complementary sensing strategy is required to characterize the ionic contribution independently of the Trop-I-specific response. Integrating an ionic-sensitive measurement with the $MoS_2$ FET response could help distinguish matrix-induced electrical variations from biomarker-dependent changes, providing a more reliable approach to Trop-I analysis in complex physiological samples.

**Chemosensors for Serum analysis**

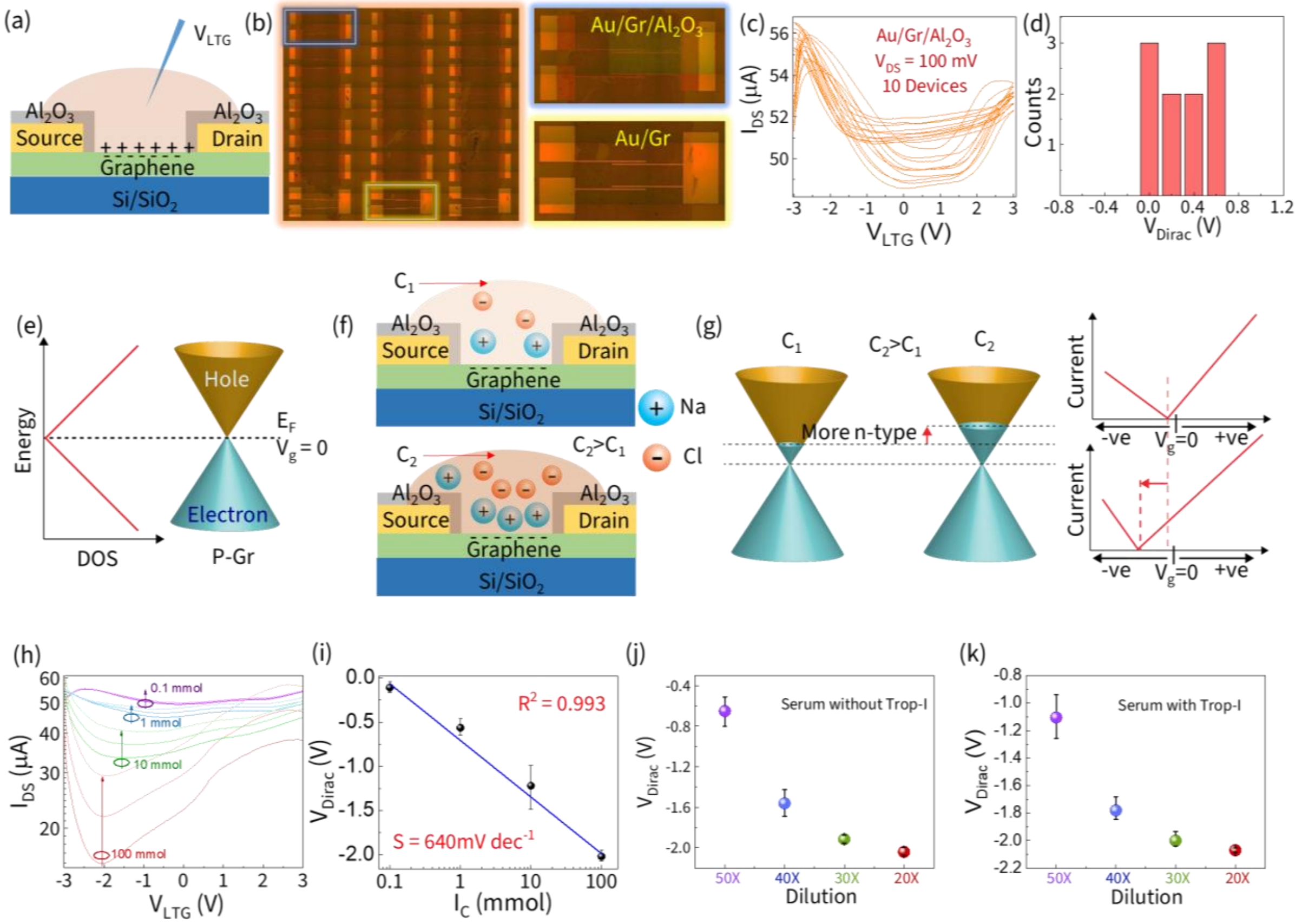

**Figure 5 Graphene-based ion-sensitive field-effect transistor (ISFET) for monitoring ionic variations in physiological samples.** (a) Schematic cross-sectional architecture of the liquid-gated graphene FET with $Al_2O_3$ passivation over the source and drain electrodes. (b) Optical images of the fabricated graphene device array and representative structures with and without $Al_2O_3$ electrode passivation. (c) Transfer characteristics of ten fabricated graphene devices measured at $V_{DS}$ =100mV. (d) Statistical distribution of the extracted $V_{Dirac}$. (e) Schematic representation of the graphene density of states and band structure, illustrating the charge-neutrality point and Fermi-level position under liquid-gating conditions. (f) Schematic illustration of electrical double-layer formation at the graphene–electrolyte interface under different ionic concentrations. (g) Proposed electrostatic doping mechanism illustrating the concentration-dependent Fermi-level modulation and corresponding negative shift of the Dirac point. (h) Transfer characteristics of the graphene ISFET measured at different ionic concentrations (0.1–100 mM). (i) Calibration curve showing the variation in Dirac-point voltage with logarithmic ionic concentration, yielding a sensitivity of 640 mV $dec^{-1}$ and a $R^2$ of 0.993. (j,k) Dirac-point voltage variation with serum dilution in the absence (j) and presence (k) of spiked Trop-I, demonstrating the dominant influence of the serum ionic environment on the graphene ISFET response.

To address the influence of the complex ionic environment of physiological samples on the $MoS_2$ FET response, a complementary graphene-based ion-sensitive field-effect transistor (ISFET) was developed for monitoring the ionic contribution. Graphene was selected as the sensing channel owing to its ambipolar transport characteristics and high sensitivity to electrostatic perturbations at its surface[28,58]. The device architecture is illustrated in **Fig. 5a**, where the electrolyte serves as the liquid gate, while $Al_2O_3$ passivation isolates the source and drain electrodes from direct electrolyte exposure, allowing the ionic interaction to occur predominantly at the exposed graphene channel[59]. The fabricated device array and representative configurations with and without electrode passivation are presented in **Fig. 5b**. The electrical characteristics of the fabricated graphene devices were evaluated to establish device reproducibility. Transfer characteristics of ten devices measured at a constant drain bias of $V_{DS}$ =100 mV are presented in **Fig. 5c**, while the corresponding statistical distribution of the extracted Dirac-point voltage is shown in Figure 5d. The Dirac point, corresponding to the charge-neutrality condition at which graphene exhibits minimum conductivity, was observed within the gate-voltage range of approximately 0–0.5 V. Although the density of states theoretically approaches zero at the Dirac point in ideal graphene, practical CVD-grown devices exhibit finite residual conductivity owing to substrate-induced doping, structural imperfections, adsorbed species, and charge inhomogeneity[60]. The ambipolar electronic structure of graphene and the associated Fermi-level modulation are illustrated in **Fig 5e** and further discussed in **Supporting Information 9**. The structural and Raman characterization of the CVD-grown graphene film, including its defect-related characteristics, is provided in **Supporting Information 10**. The ionic-sensing mechanism relies on the formation of an electrical double layer (EDL) at the graphene–electrolyte interface. Upon electrolyte exposure, ionic species accumulate near the graphene surface, inducing electrostatic doping and modifying the Fermi level[58,61] of the channel, as schematically illustrated in **Fig. 5f-g**. Changes in ionic concentration consequently alter the interfacial electrostatic potential and shift the Dirac-point voltage.[62] Under the investigated electrolyte conditions, increasing ionic concentration produces a systematic negative shift of the Dirac point, consistent with enhanced electron doping of the graphene channel. To quantify the ionic response, the transfer characteristics were recorded over an electrolyte concentration range of 0.1–100 mM (**Fig. 5h**). The extracted Dirac-point voltage exhibits a strong concentration-dependent response, with an approximately linear relationship against the logarithm of ionic concentration (**Fig. 5i**). The calibration demonstrates a coefficient of determination ($R^2$) of 0.993 and an ionic sensitivity of 640 mV $dec^{-1}$, indicating the capability of the graphene ISFET to resolve variations in ionic concentration over the investigated range[27,62]. Additional measurements addressing device repeatability, electrical leakage, recovery behaviour, and responses under different physiological ionic conditions are presented in **Supporting Information 11**.

Finally, the applicability of the graphene ISFET was investigated using human serum samples at different dilution levels, both in the absence and presence of spiked Trop-I (**Fig. 5j-k**). The sensor exhibited a pronounced dependence on serum concentration, while measurements with and without Trop-I showed similar dilution-dependent trends. This observation indicates that the graphene sensor response is predominantly governed by the ionic composition of the serum matrix rather than the specific Trop-I concentration[63]. At higher serum fractions, the device exhibited reduced recovery and a limited electrical tuning response, highlighting the influence of the concentrated physiological matrix on the electrolyte-gated interface. Dilution therefore provides a practical approach to reducing the ionic contribution and improving the measurement conditions[21,64]. Importantly, the similar responses observed in serum with and without Trop-I establish the graphene ISFET as a complementary ionic-sensitive platform rather than a selective Trop-I sensor. Integrating its ionic-response information with the $MoS_2$ FET measurements offers a potential strategy to distinguish matrix-induced electrical variations from biomarker-dependent responses, improving the interpretation of Trop-I detection in complex physiological samples[63,65].

## Duel sensors signal rectification

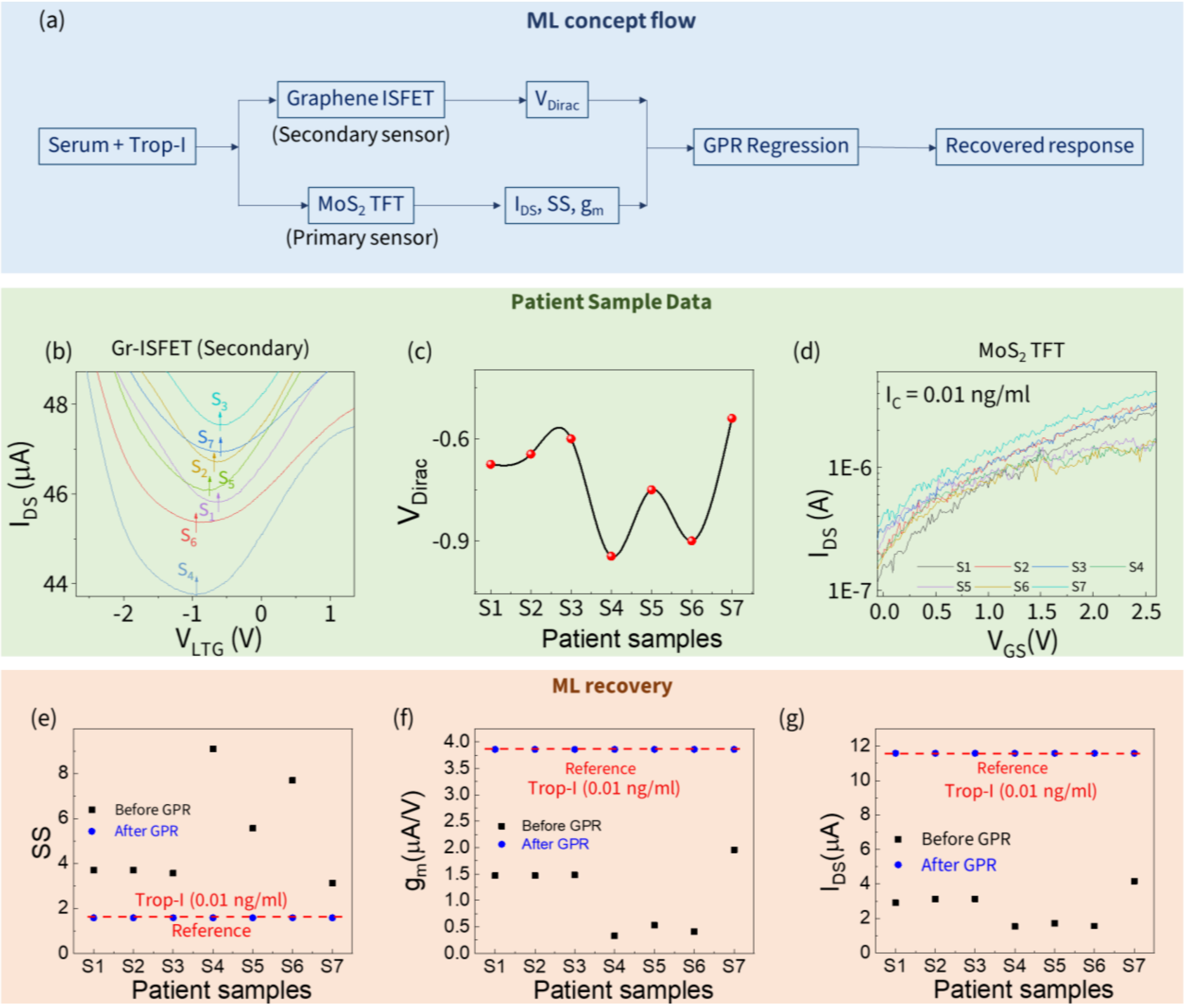

**Figure 6 GPR-assisted compensation of ionic interference in human samples**. (a) Schematic of the proposed sensing and GPR regression framework integrating $V_{Dirac}$ from the Gr-ISFET with $I_{DS}$, SS, and $g_m$ from the $MoS_2$ FET. (b) Transfer characteristics of the Gr-ISFET measured using seven human

samples (S1–S7). (c) Corresponding $V_{Dirac}$ values. (d) Transfer characteristics of the $MoS_2$ FET measured at a fixed Trop-I concentration of 0.01 ng $mL^{-1}$ across the seven human samples. (e–g) Comparison of the original responses (black squares), GPR-estimated responses (blue circles), and buffer reference values (red dashed lines) for (e) SS, (f) $g_m$, and (g) $I_{DS}$, respectively.

Having established the ionic sensitivity of the graphene ISFET, the complementary sensing platforms were integrated to investigate the feasibility of compensating serum-induced interference in the $MoS_2$ FET response. The proposed machine-learning-assisted concept is illustrated in **Fig. 6a**, where the graphene ISFET provides ionic information through the Dirac-point voltage ($V_{Dirac}$), while the $MoS_2$ FET provides biomarker-sensitive electrical information through ($I_{ON}$), SS, and $g_m$[21,63]. These complementary parameters were incorporated into separate Gaussian process regression (GPR) models[66] to estimate the corresponding buffer-equivalent transistor responses. The electrical responses obtained from seven human samples (S1–S7) are presented in **Fig. 6b–d**. The graphene ISFET transfer characteristics (**Fig. 6b**) exhibit sample-dependent variations in their Dirac-point positions, as summarized in **Fig. 6c**, indicating differences in the ionic environment among the investigated serum samples. Correspondingly, the $MoS_2$ FET transfer characteristics measured at a fixed Trop-I concentration of 0.01 ng $mL^{-1}$ (**Fig. 6d**) exhibit variations in their electrical response across the humansamples. These observations demonstrate that the serum matrix introduces sample-dependent ionic effects that can influence the electrical readout of the primary $MoS_2$ biosensor. To account for these variations, three independent GPR models were employed to estimate the buffer-equivalent values of SS, $g_m$, and $I_{ON}$, using the corresponding graphene ISFET measurements. The resulting GPR estimates are presented in **Fig. 6e–g**, respectively. The original human-sample responses (black symbols) exhibit considerable deviations from the buffer reference values (red dashed lines). Following GPR fitting, the estimated responses (blue symbols) approach the corresponding buffer references, indicating the feasibility of using complementary ionic-sensing information to compensate for matrix-associated variations in the transistor response. The simultaneous consideration of $V_{Dirac}$ and the individual $MoS_2$ electrical parameters provides a means of accounting for sample-dependent ionic contributions while retaining the biomarker-sensitive information from the primary sensing platform. Although the present analysis is based on seven human samples and requires further validation using independent samples, it demonstrates the potential of combining graphene ISFET-based ionic monitoring with GPR-assisted electrical signal compensation. This complementary sensing strategy offers a promising approach for improving the interpretation of cardiac biomarker measurements in complex physiological matrices.

The key novelties of this work are: (i) CVD-grown $MoS_2$ and graphene-based 2D material platforms are employed to develop a physiologically compatible sensing system. (ii)

Different device geometries are investigated to understand the influence of channel dimensions and active sensing area on sensitivity. (iii) To achieve improved Trop-I detection at low concentrations, the sensing strategy is extended from resistive sensing to a three-terminal TFT platform. (iv) To address the limitations arising from the high ionic content of serum, a complementary graphene-based ISFET is introduced for ionic monitoring. (v) The graphene sensor is evaluated under different physiological conditions, identifying 50-fold serum dilution as the suitable operating condition. (vi) Finally, measurements from seven human samples using both sensing platforms are integrated through a GPR model to compensate for ionic interference and estimate the buffer-equivalent Trop-I response, improving the reliability of biomarker analysis.

## Discussion

In summary, we developed a complementary 2D-material-based sensing strategy for cardiac Trop-I detection under physiologically relevant conditions. An aptamer-functionalized monolayer $MoS_2$ FET was established as the primary sensing platform, with systematic optimization of the sensing interface, device geometry, and electrical response. Although the device exhibited selective Trop-I detection with pronounced sensitivity in the low-concentration regime, its electrical performance was significantly influenced by the ionic environment of serum. To address this challenge, a graphene-based ISFET was introduced as a secondary sensor for monitoring ionic variations through Dirac-point modulation, exhibiting a sensitivity of 640 mV $dec^{-1}$ with an $R^2$ of 0.993. Finally, GPR models were employed to correlate the ionic information with the electrical response of the $MoS_2$ FET, demonstrating preliminary compensation of serum-induced variations across seven human samples. This work highlights the potential of combining complementary 2D-material sensors with machine-learning-assisted signal processing for more reliable cardiac biomarker analysis in complex physiological matrices.

## Methods

### $MoS_2$ growth and monolayer transfer

$MoS_2$ crystals and continuous layers were grown on $SiO_2$/Si substrates using an atmospheric-pressure chemical vapour deposition (APCVD) system (Quazar Technologies, QRYSTAL-CVD-1100-4). $MoO_3$ (Sigma-Aldrich, ≥99% purity) and sulfur (Sigma-Aldrich, 99.998% purity) were used as the Mo and S precursors, respectively. $SiO_2$/Si substrates (1.2 × 1.2 $cm^2$, 300 nm $SiO_2$) were sequentially ultrasonicated in acetone, DI water, and IPA for 15 min each. For growth, 2.3 mg of $MoO_3$ and 230 mg of sulfur were placed in the 8th and 4th heating zones, respectively, with the substrate positioned face-down above the $MoO_3$ precursor. After five successive pumping and purging cycles, the system was maintained under continuous Ar flow. The $MoO_3$ and sulfur zones were initially stabilized at 150°C and 50°C, respectively. The $MoO_3$ temperature was subsequently ramped to 725°C at 15°C $min^{-1}$, followed by heating the sulfur zone to 400°C. The $MoO_3$ temperature was then increased to 750°C to initiate $MoS_2$ growth, which was maintained for 20 min. After growth, the system was cooled under Ar flow, initially at 10 sccm to 500°C and subsequently at 200 sccm to room temperature (~24°C).

The as-grown monolayer $MoS_2$ was transferred onto the $Al_2O_3$ gate dielectric using a PMMA-assisted wet-transfer process. A 4 wt% PMMA solution was prepared by dissolving PMMA powder ($(C_5H_8O_2,)n$, TCI) in anisole ($C_7H_8O$, SRL, purity 99%) under continuous stirring for 12 h, followed by filtration through a solvent-compatible syringe filter to remove undissolved residues and particulate impurities. The prepared solution was stored under cool conditions until use. Subsequently, the PMMA solution was spin-coated onto the $MoS_2$-coated $SiO_2$/Si substrate using a two-step spinning process at 500 and 1500 rpm, with a total spinning duration of 2 min. The coated substrate was then maintained undisturbed for 24 h to facilitate solvent evaporation and stabilization of the PMMA support layer. The PMMA-supported $MoS_2$ was released from the growth substrate by immersion in 0.5 M NaOH solution. The released PMMA/$MoS_2$ stack was subsequently transferred through three successive DI water baths, each maintained for 1 h, to remove residual NaOH. Finally, the PMMA-supported $MoS_2$ was transferred onto the target $Al_2O_3$ substrate, followed by removal of the PMMA support layer using acetone to obtain the transferred $MoS_2$ channel.

**Fabrication of the IDE**

The interdigitated electrodes (IDEs) were fabricated using a Microwriter ML3 maskless lithography system under cleanroom conditions. Unlike conventional photolithography that requires a physical photomask, the maskless lithography system employs digitally defined patterns, providing greater flexibility in electrode design and geometry optimization. Initially, the $MoS_2$-coated substrates were cleaned sequentially with acetone, isopropyl alcohol (IPA), and deionized (DI) water, followed by drying under a nitrogen stream. Subsequently, AZ5214 photoresist (2-methoxy-1-methylethyl acetate-based formulation, Merck) was deposited onto the substrate by spin coating using two rotational speeds of 1000 and 5000 rpm for a total duration of 2 min. The coated substrates were then soft-baked at 90°C for 50 s. The IDE patterns were transferred onto the photoresist using UV exposure through the ML3 maskless lithography system, followed by development in AZ 326 MIF developer to remove the exposed photoresist regions. Metal electrodes consisting of Ti (6 nm)/Au (20 nm) were subsequently deposited by electron-beam evaporation. Finally, the unwanted metal and residual photoresist were removed by lift-off in N-methyl-2-pyrrolidone (NMP; $C_5H_9NO$, Sigma-Aldrich, 99.5% purity) at 70°C. The fabricated substrates were then sequentially rinsed with IPA and DI water and dried under a nitrogen stream.

### Fabrication of the $MoS_2$ transistor (Primary sensor)

The fabrication procedure of the $MoS_2$ FET is schematically illustrated in Supporting Information 7. A heavily doped $n^{++}$-Si substrate was used as both the device substrate and the global back-gate electrode. A 30 nm $Al_2O_3$ gate dielectric was deposited by atomic layer deposition (ALD) using trimethylaluminium (TMA) and $H_2O$ as the aluminium and oxygen precursors, respectively, as detailed in Supporting Information 6. Subsequently, the as-grown monolayer $MoS_2$ was transferred onto the $Al_2O_3$ surface, followed by source/drain electrode fabrication using photolithography and electron-beam evaporation. After metal deposition, the $Al_2O_3$ layer on the backside of the Si substrate was selectively removed by floating the substrate in buffered oxide etchant (BOE (Ammonium Fluoride (NH4F, SRL, Purity 98%), HF (Sigma Aldrich 48%)) for 30 s to expose the conductive Si surface for back-gate electrical contact. The sample was then rinsed with DI water, followed by photoresist removal using NMP and subsequent IPA and DI water cleaning. For mechanical support and stable electrical connection, the fabricated device was mounted on a glass substrate using copper tape. Electrical contact between the exposed Si backside and copper tape was established using silver conductive paste, which was allowed to dry at room temperature for 12 h. The completed devices were subsequently used for electrical characterization and biosensing measurements.

### Fabrication of the Graphene ISFET (Secondary sensor)

The fabrication procedure of the Gr-ISFET is schematically illustrated in Supporting Information (Figure S8). The IDE lectrodes were fabricated using the same photolithographic process described previously. Large-area graphene was grown on Cu foil using an AIXTRON Black Magic (PECVD/LPCVD) system operated in LPCVD mode, with acetylene ($C_2H_2$) as the carbon precursor at 1020°C. For device fabrication, 1 × 1 $inch^2$ graphene-coated Cu foil was used as the transfer source. The graphene was coated with a PMMA support layer using the previously described coating procedure. Subsequently, the underlying Cu foil was etched using 0.1 M ammonium persulfate solution, leaving the PMMA-supported graphene membrane floating on the etchant. The released membrane was sequentially transferred through three DI water baths to remove residual etchant and ionic contaminants. Prior to graphene transfer, the electrical contact pads of the pre-patterned IDE substrate were protected using photoresist, while the active channel region was left exposed for graphene integration. The PMMA-supported graphene membrane was then transferred onto the substrate by a wet-transfer process. The transferred structure was sequentially heated at 50 and 70°C to facilitate solvent evaporation and improve graphene adhesion to the $SiO_2$ substrate. Subsequently, the PMMA support layer was removed using acetone, followed by removal of the patterned photoresist through a lift-off process, leaving graphene confined to the active sensing region while maintaining accessible electrical contact pads. Finally, a patterned 50 nm $Al_2O_3$ passivation layer was deposited by electron-beam evaporation to electrically isolate the Au electrodes from direct electrolyte exposure, while retaining the exposed graphene channel for ionic sensing.

**Human sample preparation**

Whole blood was collected in sterile blood collection tubes without anticoagulant. The collected blood was centrifuged at 5000 g for 20 min at 4 °C after being allowed to clot at room temperature for about an hour. Supernatant (serum) was collected and put into a sterile microcentrifuge tube following centrifugation. To reduce the number of freeze-thaw cycles, the obtained serum samples were aliquoted and kept at -20 °C until further examination. The samples used for the study were collected under IEC-10/2026-JULY-02 with approval from the Institutional Ethics Committee (IEC) of INST, Mohali, India.

**Material and Electrical characterization (TEM, Raman, AFM, GIXRD)**

The structural, crystallographic, and morphological characteristics of the as-grown and transferred $MoS_2$ and graphene were investigated using complementary characterization techniques. High-resolution transmission electron microscopy (HR-TEM, JEOL JEM-2100) operated at an accelerating voltage of 200 kV was employed to examine the crystal structure and lattice characteristics of the 2D materials. Confocal Raman spectroscopy (WITec alpha300R) was performed under ambient conditions using a 532 nm Nd laser excitation source to evaluate the vibrational characteristics, crystalline quality, and layer-dependent properties. Atomic force microscopy (AFM; Nanonics MultiView 2000) measurements were conducted in tapping mode under ambient conditions to investigate surface morphology, thickness, and roughness of the 2D materials. AFM was additionally employed to examine the surface topography and thickness of the deposited device layers. The structural characteristics and crystalline phase of the $Al_2O_3$ gate dielectric were further investigated using grazing-incidence X-ray diffraction (GIXRD).

Electrical measurements were performed at room temperature under ambient conditions using a probe station connected to a Keysight B1500A semiconductor device analyzer. For the back-gated $MoS_2$ FET measurements, a three-probe configuration was employed, comprising two probes for the top source and drain electrodes and an additional probe for the backside $n^{++}$-Si gate contact. The B1500A analyzer was equipped with a low-power source/monitor unit (SMU, Keysight B1511B) and a high-power SMU (Keysight B1510A), which were used according to the electrical measurement requirements. The transfer and output characteristics were recorded to evaluate the transistor performance and concentration-dependent sensing response.